\documentclass[12pt,a4paper]{article}
\usepackage{graphicx}
\usepackage{amsmath}
\usepackage{subfigure}

\begin{document}

\begin{center}
{\bf \Large Stable states of systems of bistable magnetostrictive wires against applied field, applied stress and spatial geometry
}\\[5mm]

{
\large P. Gawro\'nski$^{*1}$, A. Chizhik$^{**2}$, J. M. Blanco$^3$ and K. Ku{\l}akowski$^{***1}$  \\[3mm]

\em {$^1$ Faculty of Physics and Applied Computer Science, AGH University of Science and Technology, al. Mickiewicza 30, 30-059 Krak\'ow, Poland\\
$^2$ Departamento Fisica de Materiales, Facultad de Quimica, UPV, 1072, 20080 San Sebastian, Spain\\
$^3$ Departamento Fisica Aplicada I, EUPDS, UPV/EHU, Plaza Europa, 1, 20018 San Sebastian, Spain\\
$^*$E-mail: gawron@newton.ftj.agh.edu.pl\\
$^{**}$E-mail: wuxchcha@ehu.es\\
$^{***}$E-mail: kulakowski@novell.ftj.agh.edu.pl\\

\today
}
}
\end{center}

\begin{abstract}
Long-range magnetostatic interaction between wires strongly depends on their 
spatial position. This interaction, combined with applied tensile stress, 
influences the hysteresis loop of the system of wires through the stress 
dependence of their coercive fields. As a result, we obtain a set of stable 
magnetic states of the system, dependent on the applied field, applied stress 
and mutual positions of the wires. These states can be used to encode the 
system history.

\end{abstract}



\section{Introduction}

Amorphous magnetic wires are of interest for their actual and potential use, e. g. in sensors 
\cite{vah,vaz,moh}. As it is almost usual in micromagnetism, their properties depend on the technique 
of sample preparation and thermal treatment. Then, they are a good subject for computational science.
The problem of the stray field and/or magnetization distribution of the wire have been treated either 
by means of a purely computational 
methods, where the wire volume is divided into magnetically polarized units \cite{her,for}, 
or within the dipolar or similar model approximation \cite{pir,kac}. The latter approach is particularly 
suitable for magnetic arrays, formed as sets of parallel microwires perpendicular to the array 
\cite{vel}. Other authors used an approximation of a homogeneous interaction field \cite{sam,sav}.
As it was pointed out by Velazquez et al. \cite{pir}, the homogeneous field does not work for short 
interwire distances. Moreover, in fact the spatial distribution
of the magnetization near the wire ends remains unknown; and it is precisely in this area where 
the remagnetization process starts. As a consequence, our knowledge on the subtle process
of the switching of the magnetic moment in the wire advances with difficulties, despite an obvious 
interest motivated by technology. Recent progress in the field is due to an application of 
some phenomenological theories, and by the obtained accordance of the results with experimental 
data \cite{var}. Although this procedure improved much our understanding of the underlying physical 
mechanisms, still questions seem to be not much less valuable than answers.

Here we are interested in the bistable hysteresis loops for various orientations of the wires with respect
to each other and to the applied field. The motivation is the potential flexibility of the hysteresis loops, which we
expect to result when we modify mutual positions and orientations of the wires. On the other hand, it 
is desirable to confront the experimental data with the model calculations. Even simplified, the model
provides a reference point, and the departures between model and reality tells us where an established 
picture fails.

Our method here is to use a simplified model of the interaction, where the wires
are magnetized homogeneously and the wire diameter is set to zero. This leaves the stray field of a 
wire reduced to the field created by two magnetic charges, placed at the wire ends. The approximation
of an infinitely thin wire is justified as long as the wire-wire distance is much larger than the wire
diameter. The resulting formula is more refined than the approximation of a homogeneous field, 
and much easier to use than the micromagnetic simulation. This advantage allows to calculate the 
interaction field for any mutual position and/or orientation of the wires. 

Despite of its practical ease, there is an additional methodological argument for the assumption on the homogeneous 
distribution of the magnetization. Let us suppose the case of two parallel wires, which form two edges
of a rectangle. This system has been investigated by several authors \cite{vaz,vel,sam}. 
The experimental results on the hysteresis clearly prove that the stray field of the wires act as to 
prefer the antiparallel orientations of their magnetic moments. It is reasonable to expect that
the stray field of one wire acting at another wire should be evaluated near the end of the latter wire;
it is at this point where the remagnetization process starts. Then, every deviation from the homogeneous 
distribution of the magnetization weakens the interaction field, what makes the model uneffective. Let 
us add that in our numerical simulations reported from Section 3 we are going
to evaluate the interaction field not near the wires ends, but at the wires ends. This simplification
is expected not to change results remarkably as long as the wires are not parallel.

We are interested in two questions: i) which geometrical configurations
of the wires allow for the bistability, ii) what kinds of the hysteresis loops appear for various 
wire configurations? The former question appears as a consequence of the fact, that for some mutual 
positions and orientations of the wires, the effective (external + interaction) field causing the 
large Barkhausen jump at one wire end changes its sign at a point along of the wire; then the domain 
wall is stopped at this point and the
bistability is lost. In this case some more complex behaviour of the hysteresis loop can be expected;
for clarity we limit the subject of this text to the bistable case. The second question is a 
necessary introduction to consider magnetic properties of sets of magnetostatically interacting elements.
Such systems are of potential interest e.g. for submicrometer magnetic dots \cite{cov}. We hope that the 
answer, even simplified, will activate the search of spatial configurations of the wires which can be 
useful for other new applications.

The condition applied here can seem to be too restrictive, because in fact the bistability can still be
observed if the point at the wire where the field along the wire is zero may vanish at an increasing 
applied field before the domain 
wall reaches it. Here, however, we are not going to rely on these dynamical effects. We limit our interest 
to the cases where the bistability is observed even in the limit of small frequency, and the field amplitude
only as large as to activate the switching process. On the other hand, it is known that at field below
the order of 1 A/m the domain wall motion is much slower. However, this effect does not influence the 
shape of the hysteresis loop as long as the period 1/$f$ of the applied sinusoidal field is larger than 
the time of the remagnetization. Our discussion is limited to this case.

It is known that the wire properties depend on the applied tensile stress \cite{mit}. For the purposes of this 
work, most important agent is the stress dependence of the switching (coercive) field. We should repeat
here that the condition of the bistability considered in this text is applied for the effective field
(applied field + interaction) which triggers the domain wall motion. The enhancement of the switching field 
by the tensile stress 
demands an increase of the external field, what influences the bistability condition. Then we 
consider the tensile stress as an important and nontrivial factor which can change the system properties.

The layout of this text is as follows. In the next section the data are provided on the experimental dependence
of the switching field of the wire on the applied tensile stress. In Section 3
we outline the question on the bistability 
of interacting wires, and we ilustrate this question by an analytical solution in the simplest case of parallel wires.
In Section 4, conditions of the bistability are investigated numerically. This section contains also some 
examples of the hysteresis loops of various bistable systems, including the case with the applied tensile 
stress. Final conclusions close the text.

\section{Experiment}

In order to capture the dependence of the switching field vs the tensile stress 
the axial hysteresis loop of as-cast wire of the nominal composition 
$Fe_{77.5}B_{15}Si_{7.5}$, the diameter $125$ $\mu m$ and the length of $100$ 
$mm$ was measured by means of the conventional induction method \cite{vzq}. 
The frequency of the external sine-like field was equal to $50$ $Hz$ and the 
amplitude of was set to $40$ $A/m$. One end of a wire was fixed to the sample 
holder, while to the other one a mechanical loading was attached in order to 
apply the additional tensile stress $\sigma $ during the magnetic measurement. The 
mechanical loading was being changed from zero to $600$ $g$. Obtained values of the 
switching field $H^*(\sigma)$ are shown in Fig. 1.

\begin{figure}
\begin{center}
\resizebox{0.80\hsize}{!}{\includegraphics[angle=-90]{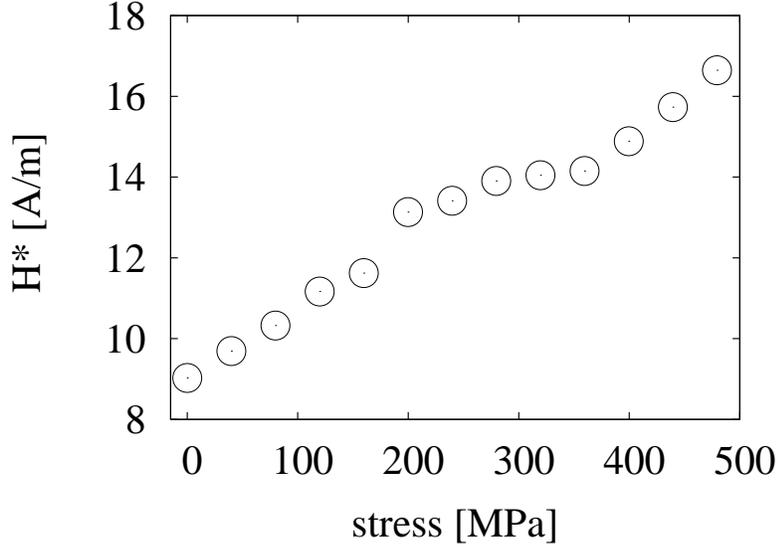}}
\end{center}
\caption{\label{exp-dat} Experimental values of the switching field $H^*$ against the applied tensile stress.}
\end{figure}

\section{The bistability of parallel wires}

The wires considered here are iron-based straight pieces of amorphous alloy, $L=2a$=10 cm long, with 
diameter of 2$R$=120 micrometers. Their magnetization is $M=0.7T/ \mu _0$=5.6$\times 10^5$ A/m. This kind of wires 
are well worked out in 
the literature \cite{che}. The magnetic pole at each end contains the magnetic charge 
$Q=\pm M\pi R^2$= $\pm $0.0063 Am. The 
switching (coercive) field is about 10 A/m. Suppose that a center of such a wire is placed at the 
coordination center, and the wire is parallel to OZ-axis. In the surrounding space, there are planes 
where the vertical component of the stray field is equal to the applied field, equal in this case to the 
switching field. If a system of two wires is to be bistable, the second wire cannot go through such a plane. 
If it does, the domain wall in the second wire is stopped.

In the (r-z) plane, the condition that the field created by one magnetic pole plus the external field
$H$ cancel with the field created by another magnetic pole is

\begin{equation}
F(z+a)=F(z-a)+ \frac{4\pi H}{Q}
\end{equation}
where
\begin{equation}
F(y)=\frac{y}{(r^2+y^2)^{3/2}} 
\end{equation}

For $H$=0, the equation can be untangled. Expressing all lengths in units of $a$, what is a half of the wire length, 
and after some simple manipulation we get the equation 

\begin{equation}
r^2 = \frac{((z+1)^2(z-1)^{2/3} - (z-1)^2(z+1)^{2/3}}{(z+1)^{2/3} - (z-1)^{2/3}}
\end{equation}

One of the consequences of this condition is that parallel 
wires of exactly the same length and very close to each other cease their bistability if they are mutually shifted
along their length. Let us consider the case when they are not shifted; then they form edges of a rectangle. 
The axial component of the stray 
field created at an end of one wire at distance $x$ from the closer end of another wire is 

\begin{equation}
H = \frac{Q}{4\pi}\frac{x}{(x^2+d^2)^{3/2}}
\end{equation}
where $d$ is the interwire distance. This component is maximal at the point $x$ where $2x^2=d^2$, and its value at the
maximum varies with $d$ approximately as $Q/(32.6d^2)$. If $d<<L$, the contribution to the stray field from other 
wire end is negligible. However, this evaluation relies on an assumption that the remagnetization starts at the 
point at the wire where the stray field from another wire has a maximum. This assumption remains not justified.
Also, the obtained values of the interaction field much excess the observed value, which is of order of 10 A/m
for the distance $d$ of 1 mm \cite{gwr}. This can mean, that in fact the switching starts at smaller value of $x$,
i.e. closer to the wire end. Anyway, the evaluation $H \propto d^{-2}$ seems to better reflect the experimental data 
of Ref. \cite{pir}, than the formula used in Ref. \cite{sam} for microwires
\begin{equation}
H_{int}=\frac{QL}{4\pi}(d^2+L^2)^{-3/2}
\end{equation}
at least for small interwire distance $d$. The latter formula can be derived as the interaction between the opposite
ends of the wires. We should add that the authors of Ref. \cite{sam} used is to test the stray field dependence on 
the wire length. The stress dependence of the point $x$ where the remagnetization starts could explain
the observed variation of the interaction field with stress \cite{gwr}. The values of the interaction
field, observed in this experiment, are as large as 500 A/m. This order of magnitude can be reproduced by
the formula $Q/(32.6d^2)$, given above. Then, the observed difference between the conventional 
and the cold-drawn wires can be discussed in terms of the position of the point, where the 
remagnetization starts.

\begin{figure}
\begin{center}
\resizebox{0.80\hsize}{!}{\includegraphics[angle=-90]{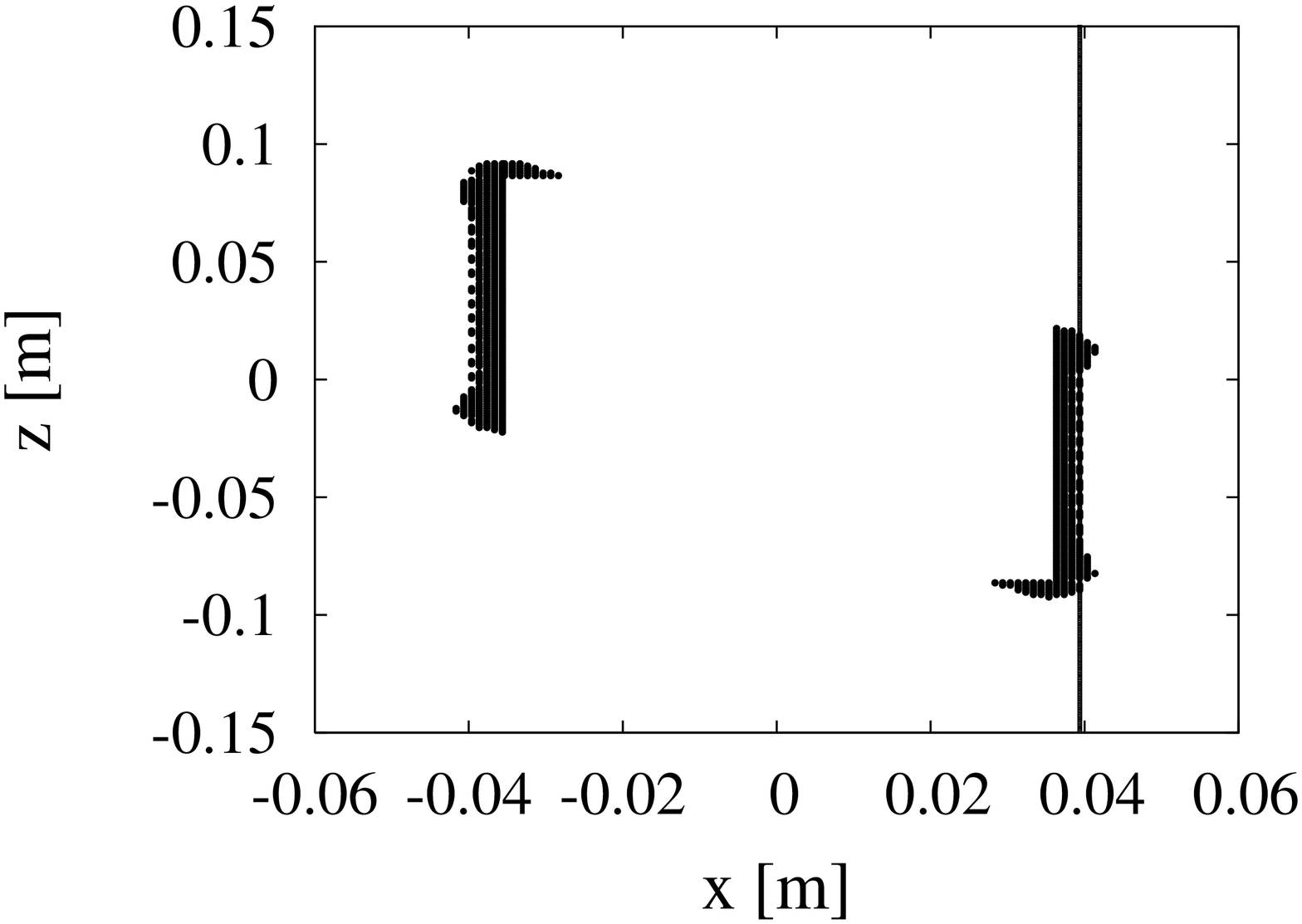}}
\resizebox{0.80\hsize}{!}{\includegraphics[angle=-90]{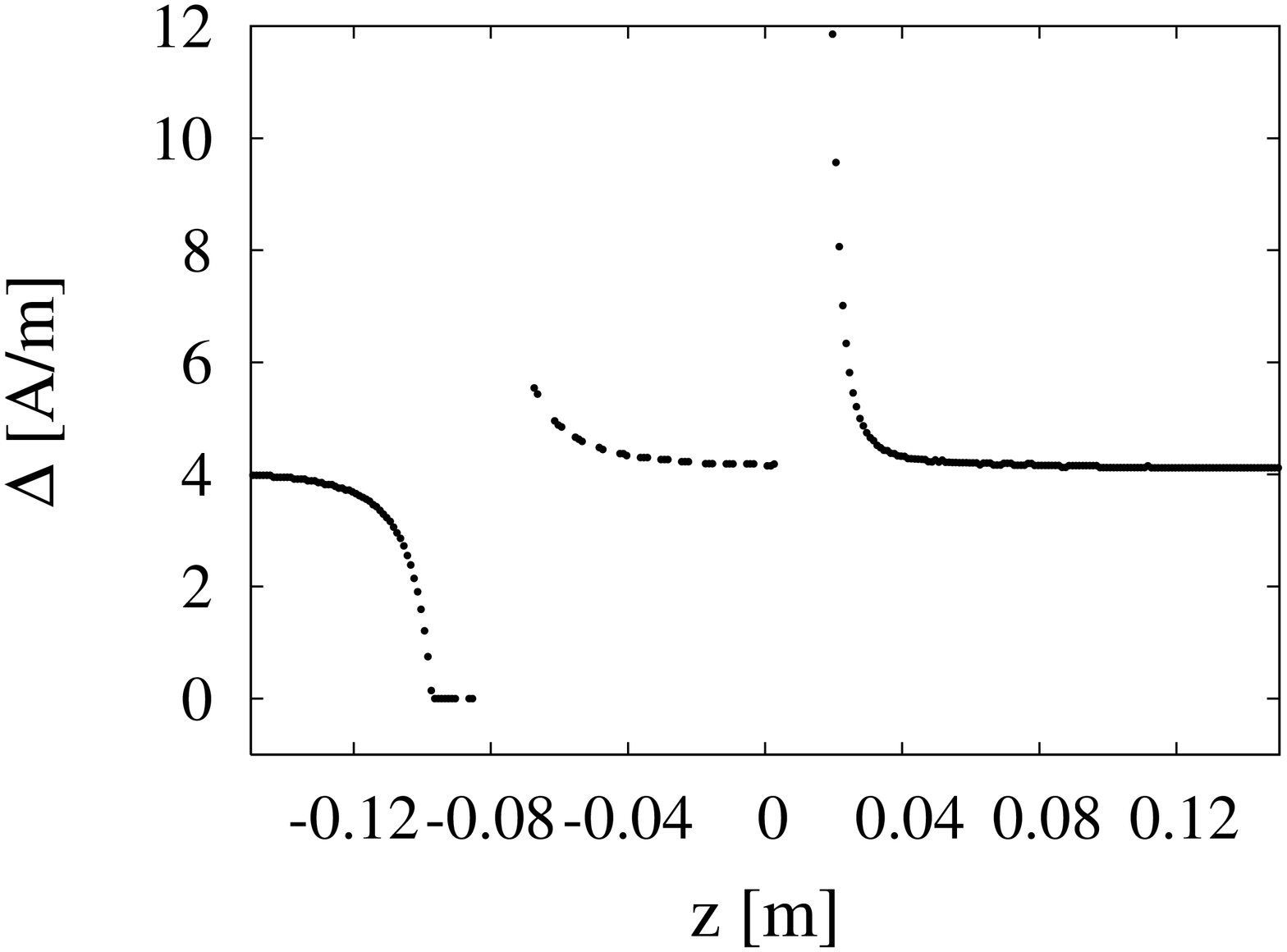}}
\end{center}
\caption{\label{fig-plot2} (a) Area where the bistability is absent. The coordinates $(x,z)$ mark the center of the wire B
with respect to the center of the wire A. The vertical solid line marks the positions of the centre of wire B for Fig. 2 b. (b) The 
horizontal part $\Delta$ of the hysteresis loop against the position of wire B.}
\end{figure}

\begin{figure}
\begin{center}
\resizebox{0.80\hsize}{!}{\includegraphics[angle=-90]{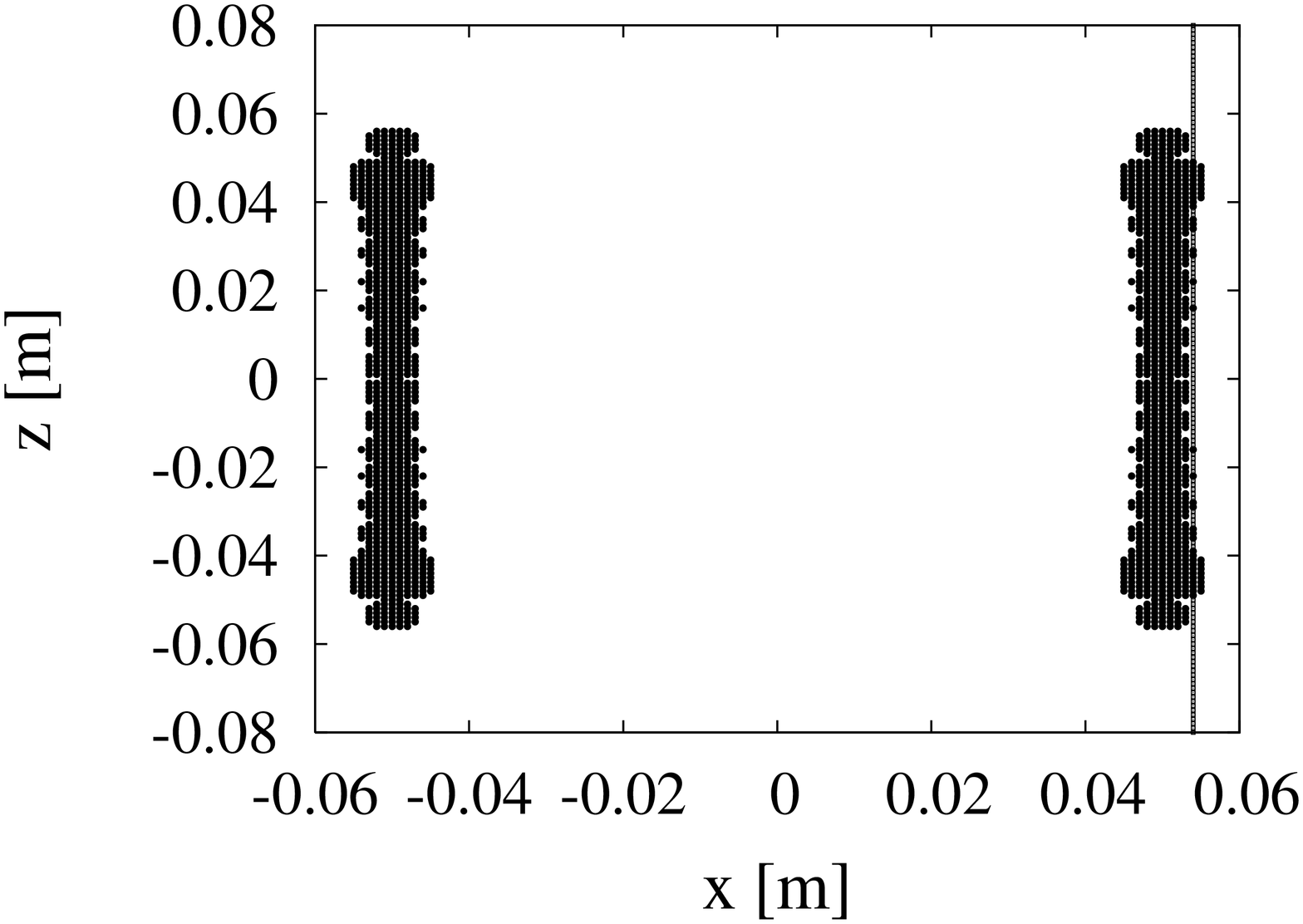}}
\resizebox{0.80\hsize}{!}{\includegraphics[angle=-90]{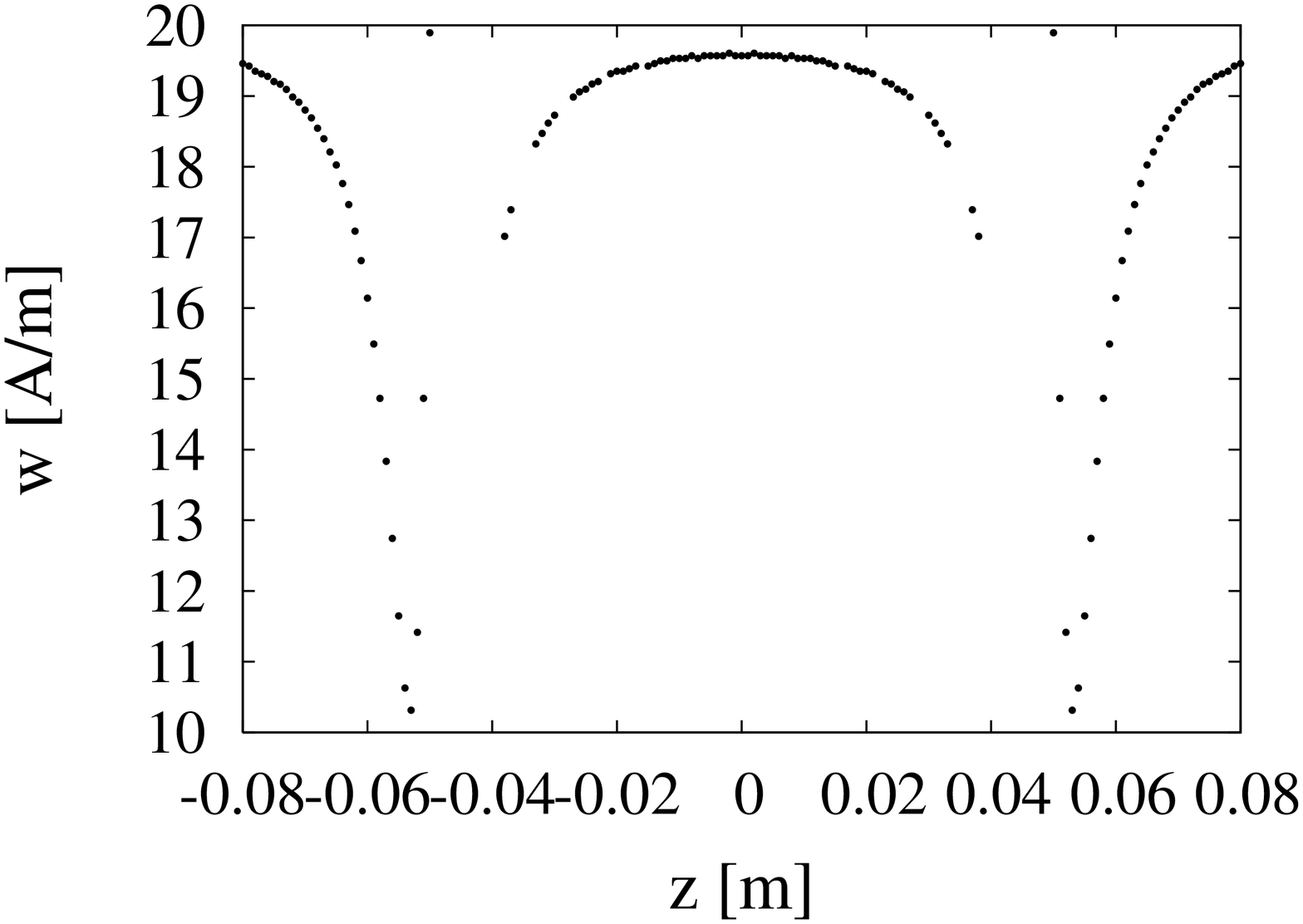}}
\resizebox{0.80\hsize}{!}{\includegraphics[angle=-90]{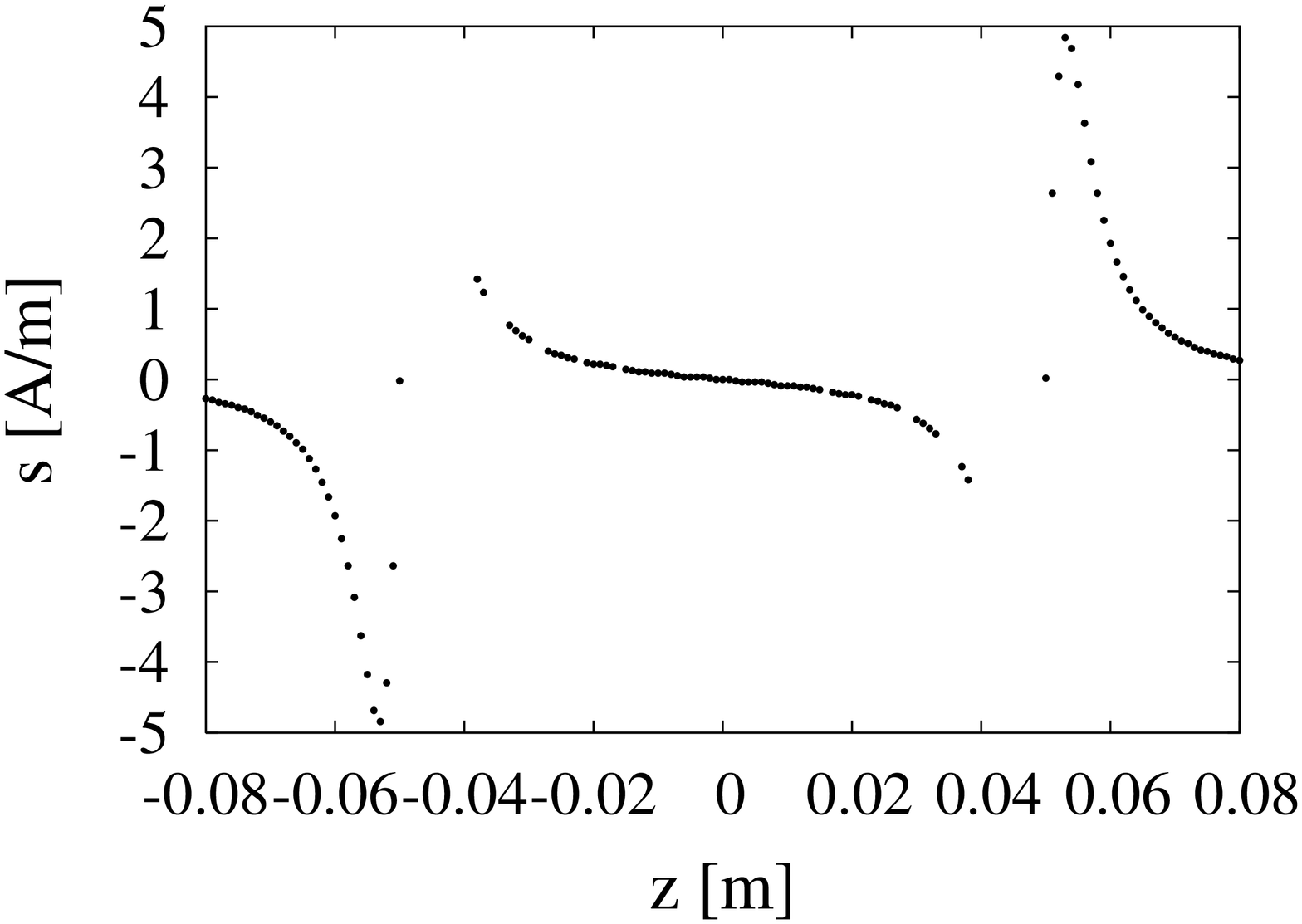}}
\end{center}
\caption{\label{fig-plot3} (a) Area where the bistability is absent. The coordinates $(x,z)$ mark the center of the wire B
with respect to the center of the wire A. The vertical solid line marks the positions of the centre of wire B for Fig. 3 b. (b,c) The 
width $w$ and shift $s$ of the centre of the hysteresis loop against the position of wire B.}
\end{figure}

\section{Conditions of the bistability - numerical solutions}

In Figs. 2-8 we show numerical solutions of the same problem for various mutual orientations of the wires. 
The value of the applied field is taken as to get the remagnetization. This means that the effective field
at one end of the wire is equal to the switching field. Here again, the question is: will the domain wall reach 
the other wire end? Marked areas in the figures give the values of the parameters, where the answer is "no".

\begin{figure}
\begin{center}
\resizebox{0.80\hsize}{!}{\includegraphics[angle=-90]{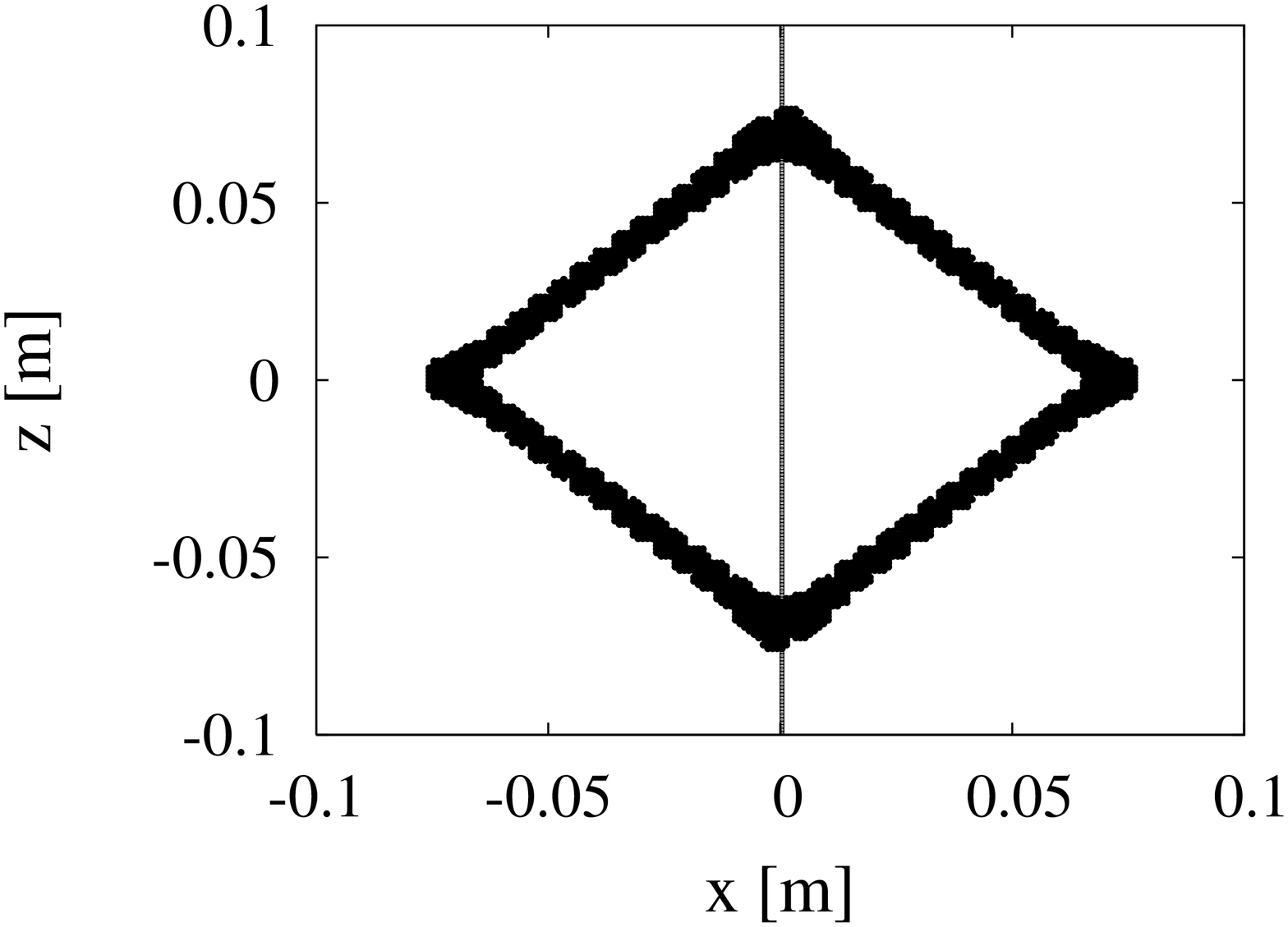}}
\resizebox{0.80\hsize}{!}{\includegraphics[angle=-90]{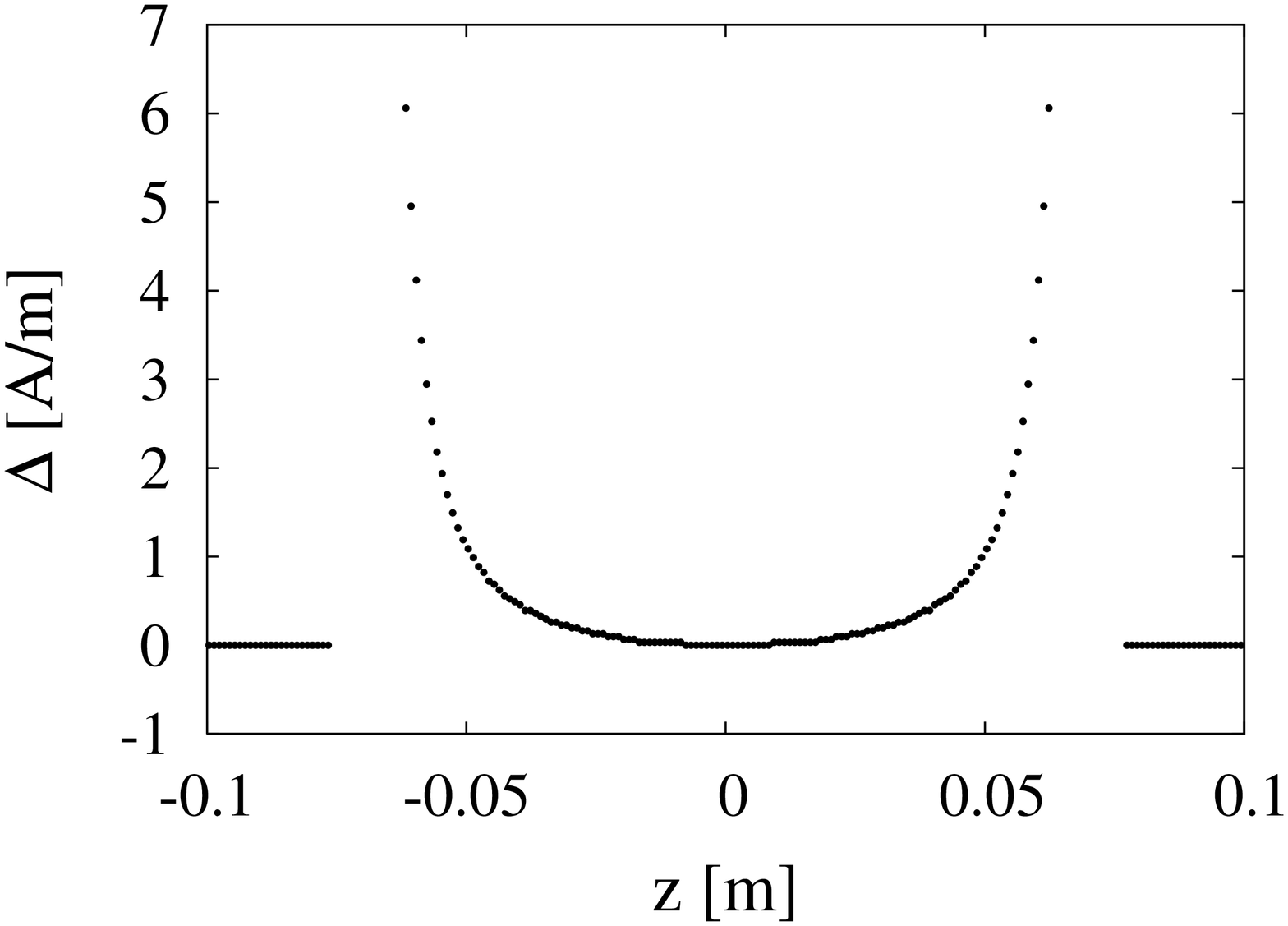}}
\end{center}
\caption{\label{fig-plot4} (a) Area where the bistability is absent. The coordinates $(x,z)$ mark the center of the wire B
with respect to the center of the wire A. The vertical solid line marks the positions of the centre of wire B for
Fig. 4 b. (b) The horizontal part $\Delta$ of the hysteresis loop against the position of wire B.}
\end{figure}

\begin{figure}
\begin{center}
\resizebox{0.80\hsize}{!}{\includegraphics[angle=-90]{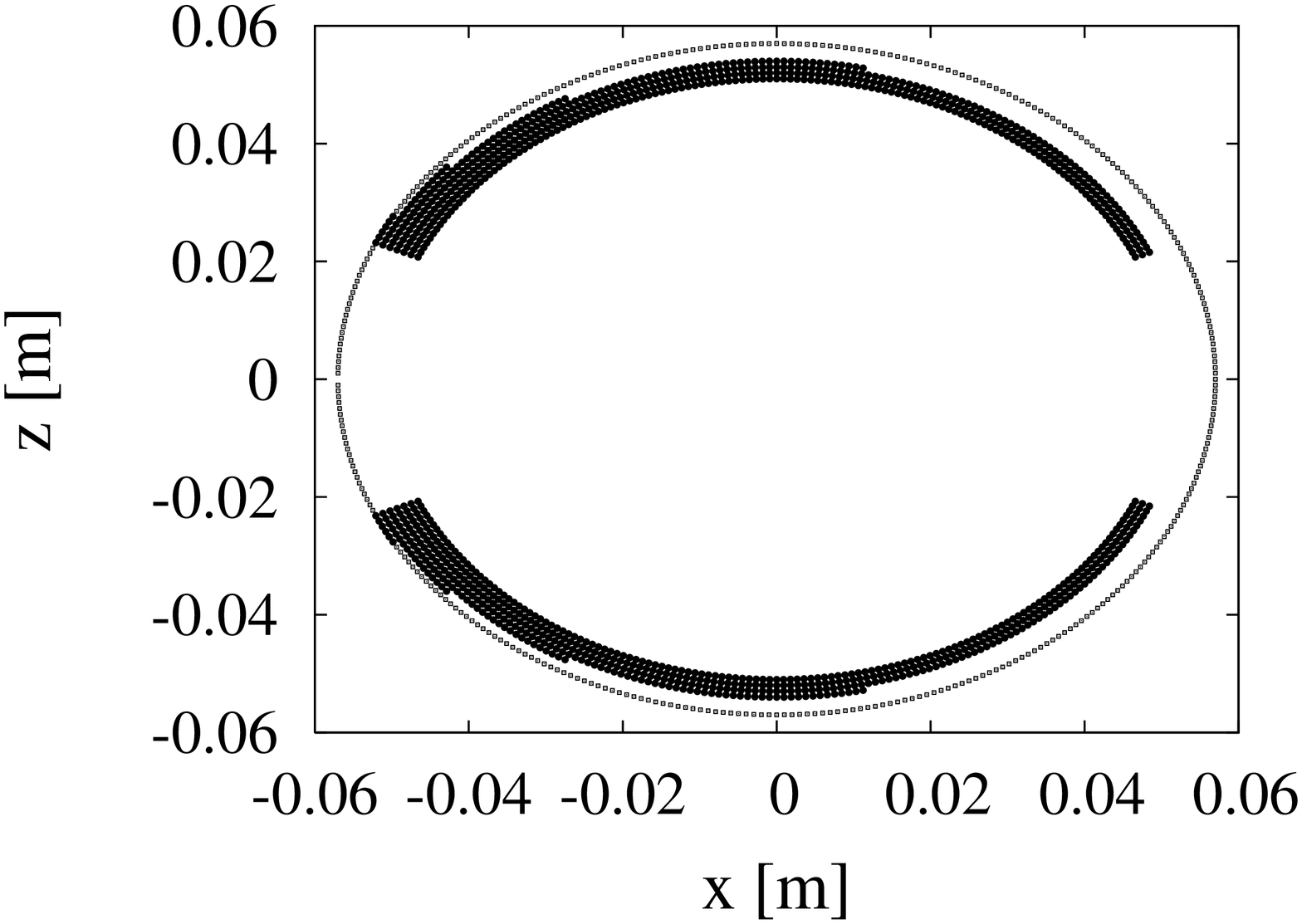}}
\resizebox{0.80\hsize}{!}{\includegraphics[angle=-90]{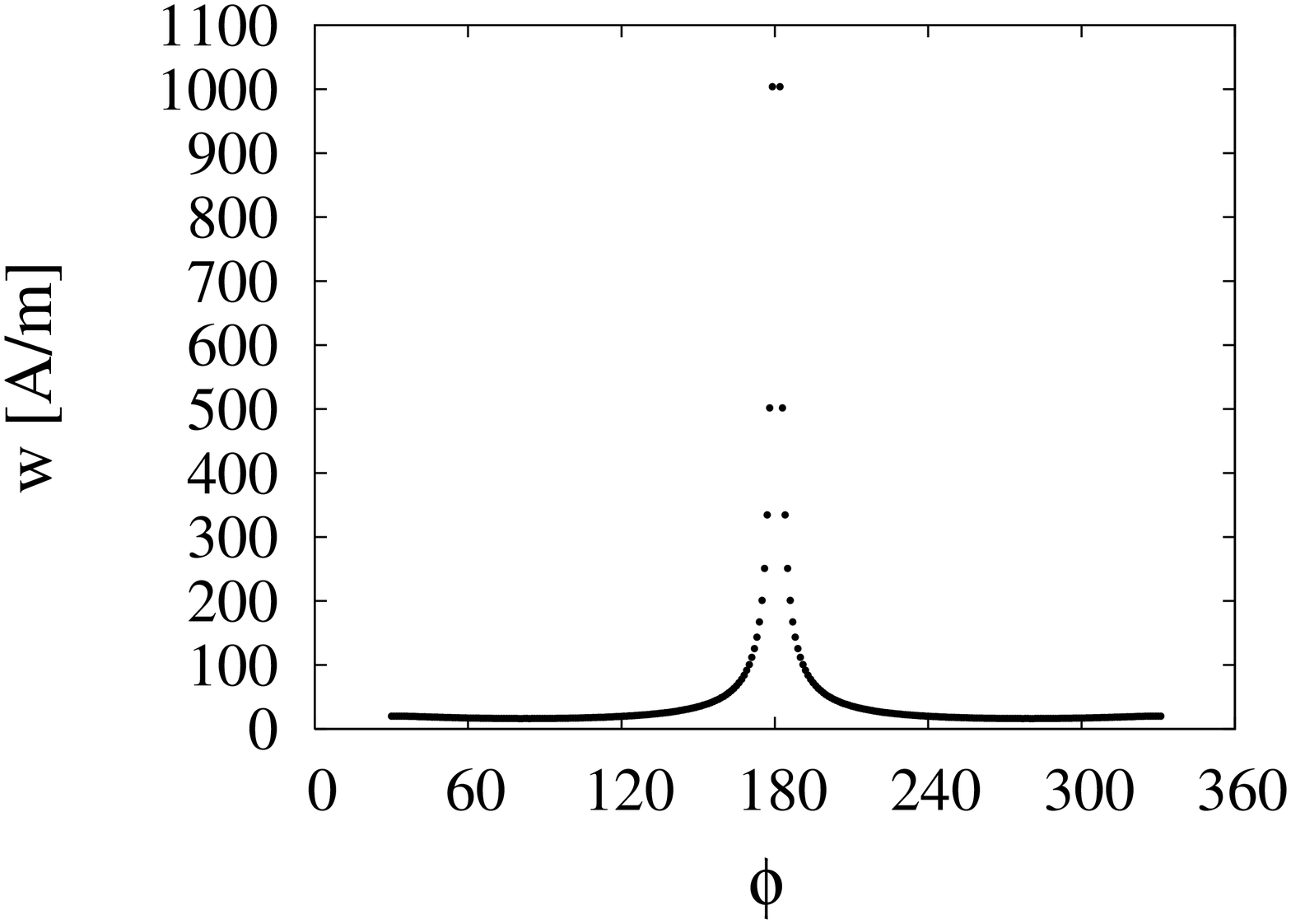}}
\resizebox{0.80\hsize}{!}{\includegraphics[angle=-90]{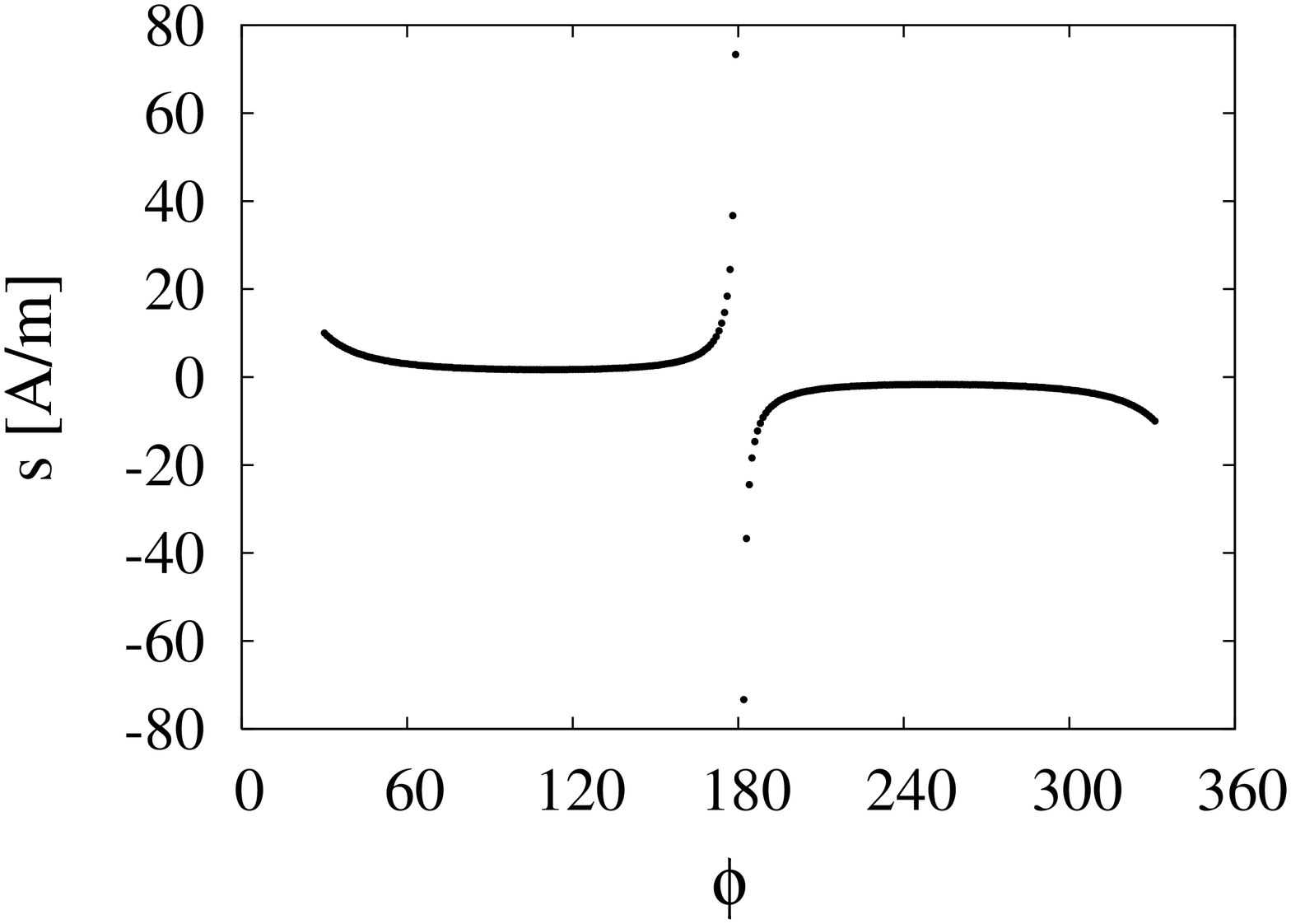}}
\end{center}
\caption{\label{fig-plot5} (a) Area where the bistability is absent. The coordinates $(x,z)$ mark the center of the wire B
with respect to the center of the wire A. The elliptical line marks the positions of the centre of wire B for Fig. 5 b. (b,c) The 
width $w$ and shift $s$ of the centre of the hysteresis loop against the position of wire B.}
\end{figure}

\begin{figure}
\begin{center}
\resizebox{0.80\hsize}{!}{\includegraphics[angle=-90]{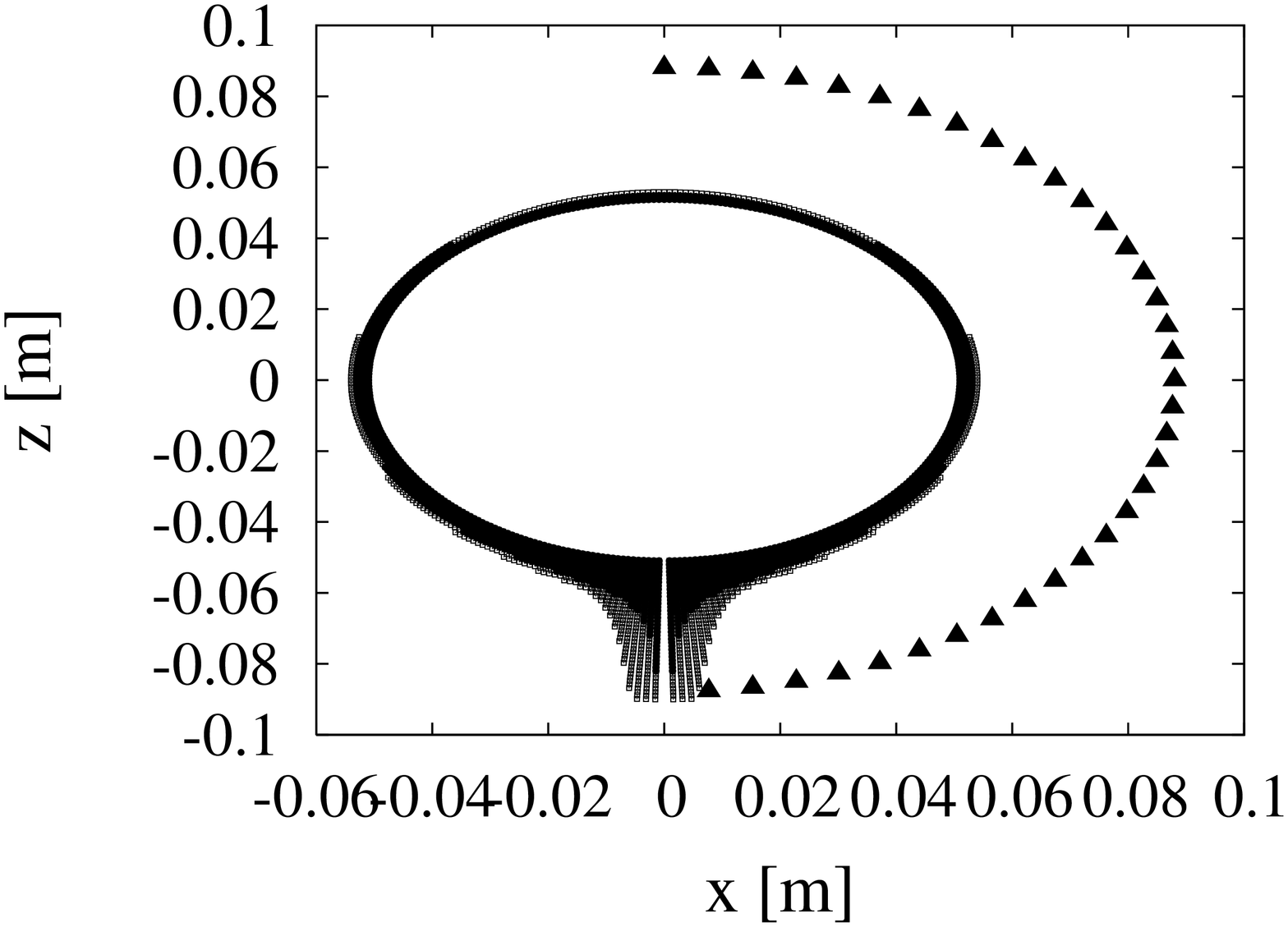}}
\resizebox{0.80\hsize}{!}{\includegraphics[angle=-90]{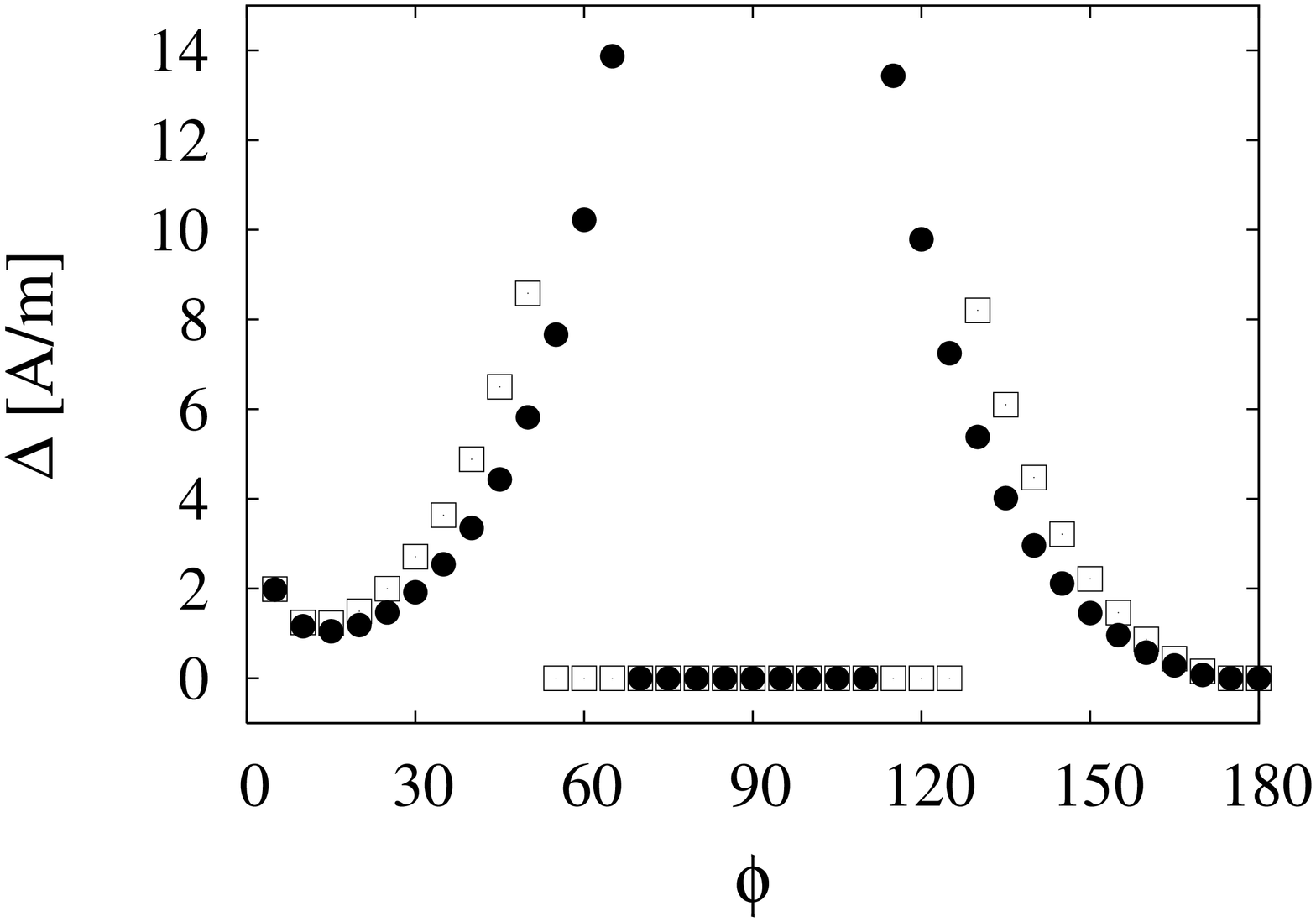}}
\resizebox{0.80\hsize}{!}{\includegraphics[angle=-90]{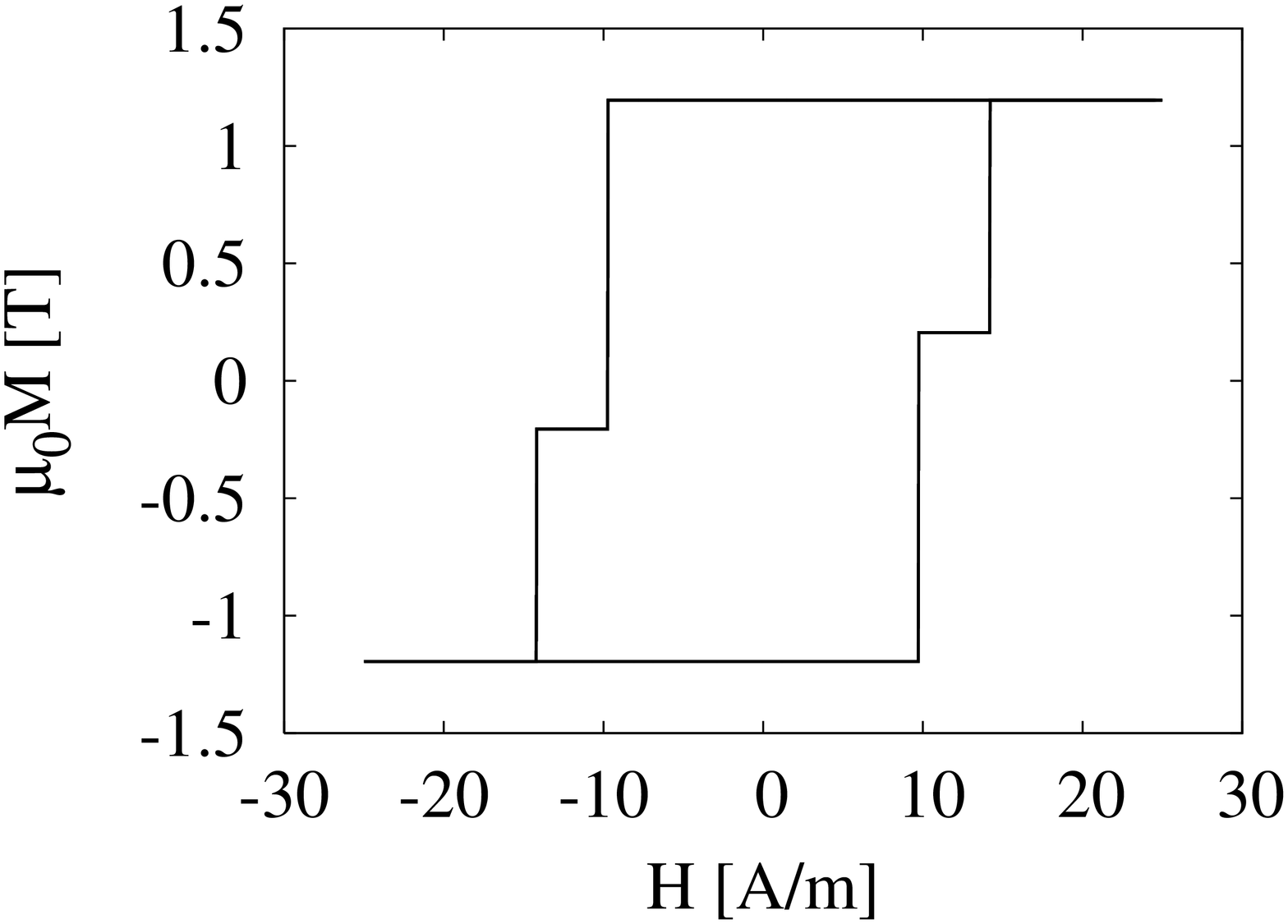}}
\end{center}
\caption{\label{fig-plot6} (a) Area where the bistability is absent: full dark (with stress) and grey (without stress). 
The coordinates $(x,z)$ mark the center of the wire B
with respect to the center of the wire A. The line of triangles marks the positions of the centre of wire B for Fig. 6 b. 
(b) The horizontal part $\Delta$ of the hysteresis loop against the position of wire B. Two sets of points mark
the results for the case with (empty squares) and without (full circles) tensile stress. (c) Example of the 
hysteresis loop with well-visible horizontal part $\Delta$.}
\end{figure}

\begin{figure}
\begin{center}
\resizebox{0.80\hsize}{!}{\includegraphics[angle=-90]{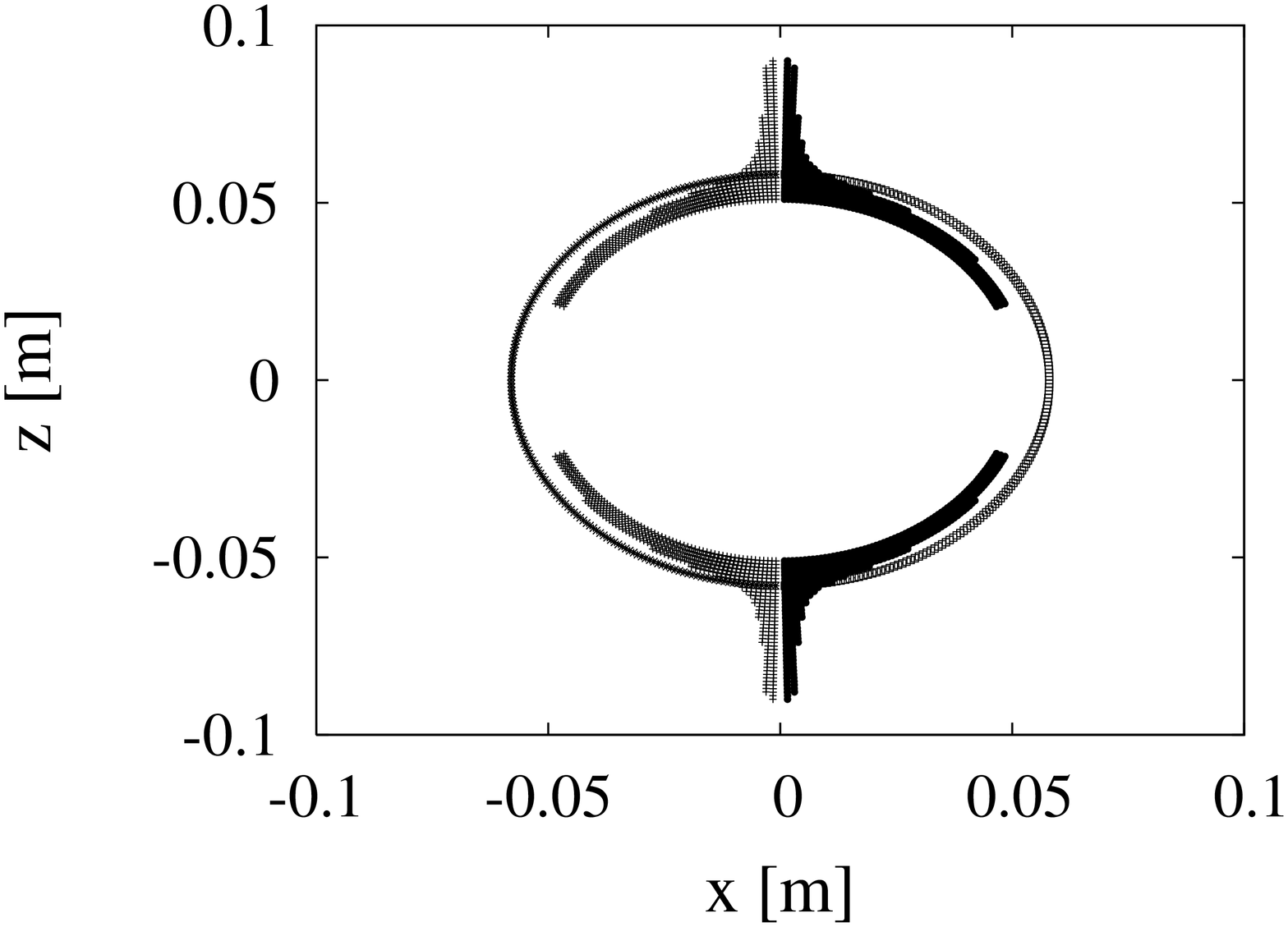}}
\resizebox{0.80\hsize}{!}{\includegraphics[angle=-90]{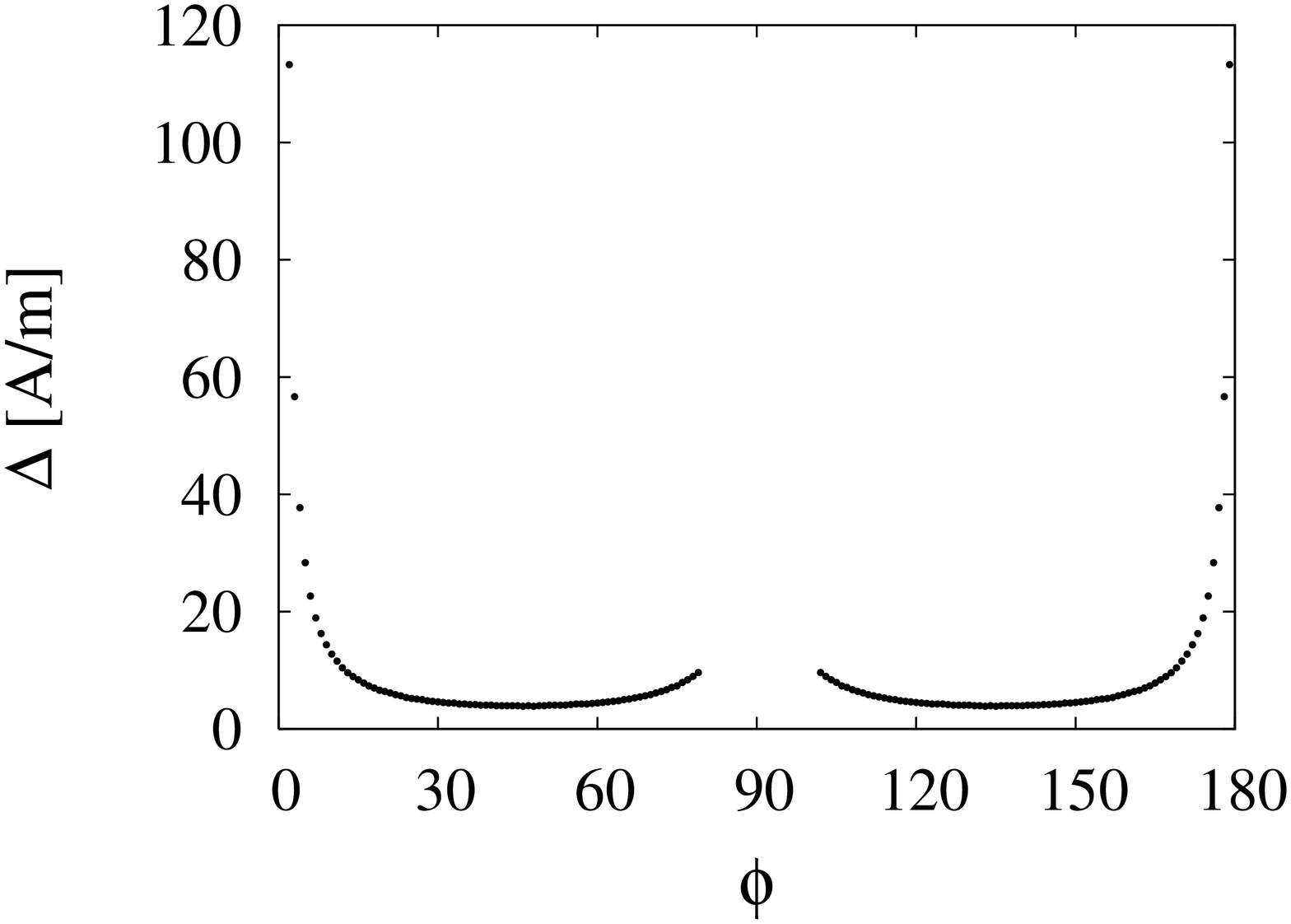}}
\end{center}
\caption{\label{fig-plot7} (a) Area where the bistability is absent. The coordinates $(x,z)$ mark the center of wires A, B
with respect to the center of coordinates. The elliptical line marks the positions of the centres of the wires for 
Fig. 7 b. (b) The 
horizontal part $\Delta$ of the hysteresis loop against the position of wire B.}
\end{figure}

\begin{figure}
\begin{center}
\resizebox{0.80\hsize}{!}{\includegraphics[angle=-90]{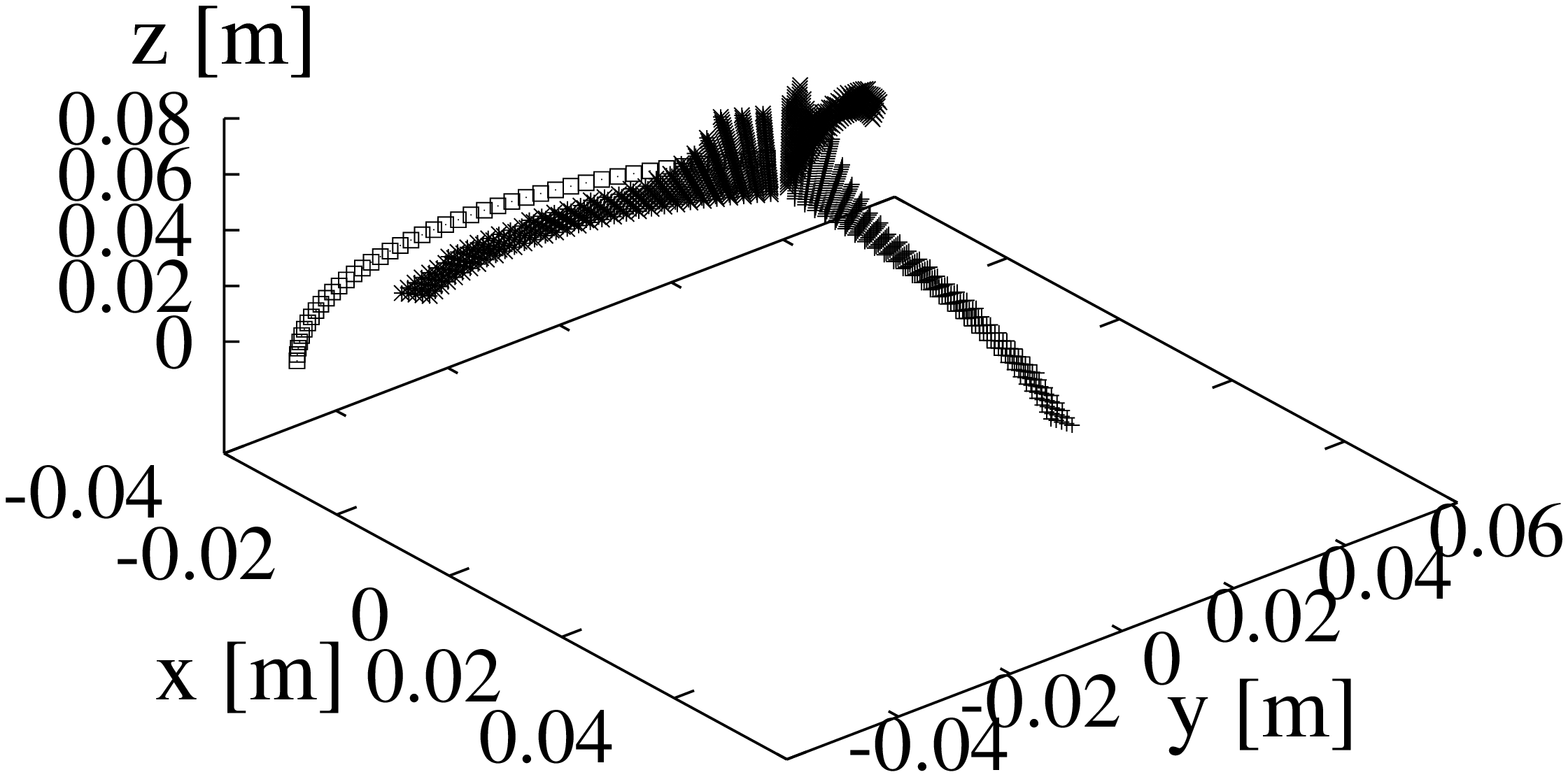}}
\resizebox{0.80\hsize}{!}{\includegraphics[angle=-90]{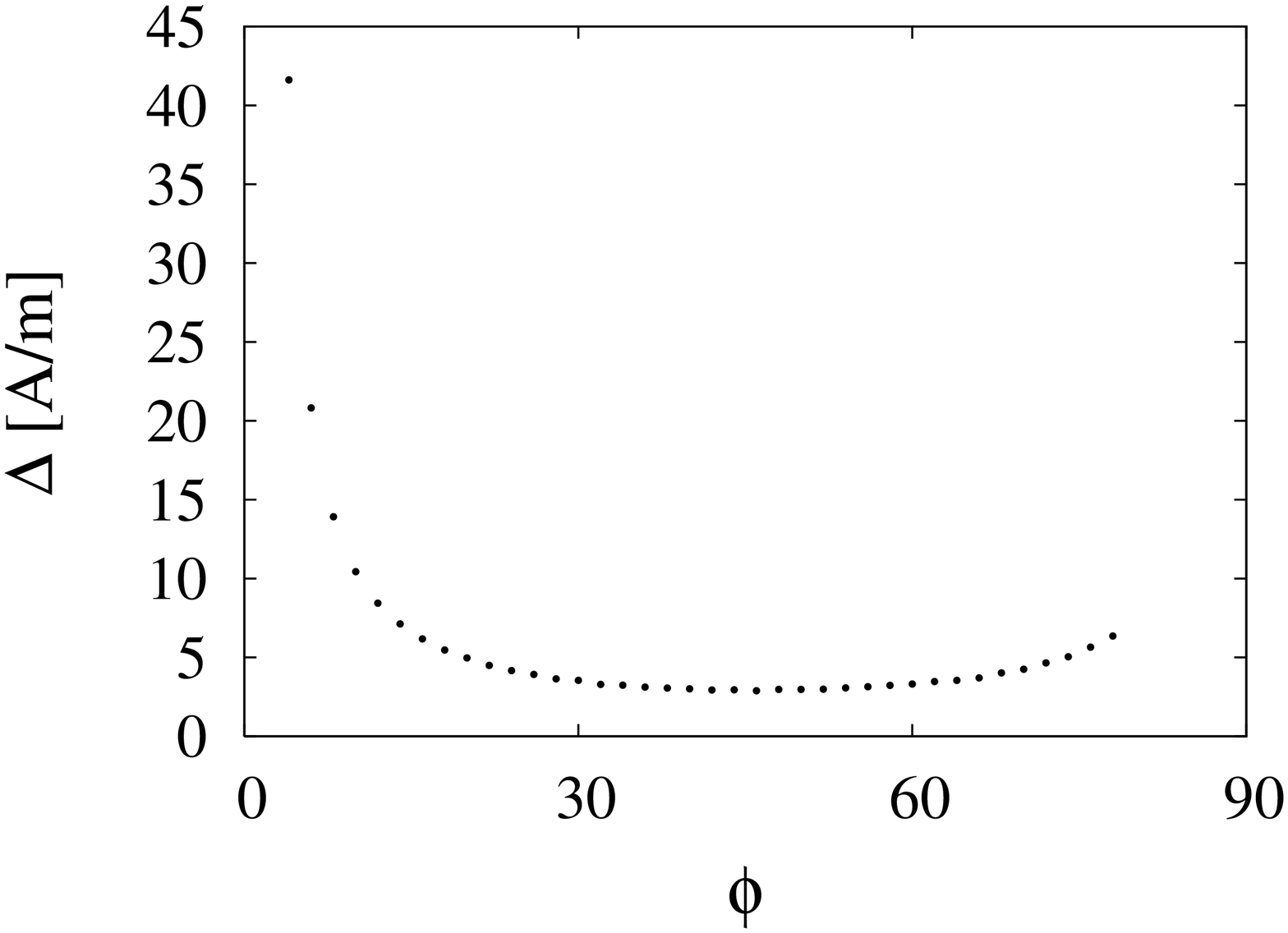}}
\resizebox{0.80\hsize}{!}{\includegraphics[angle=-90]{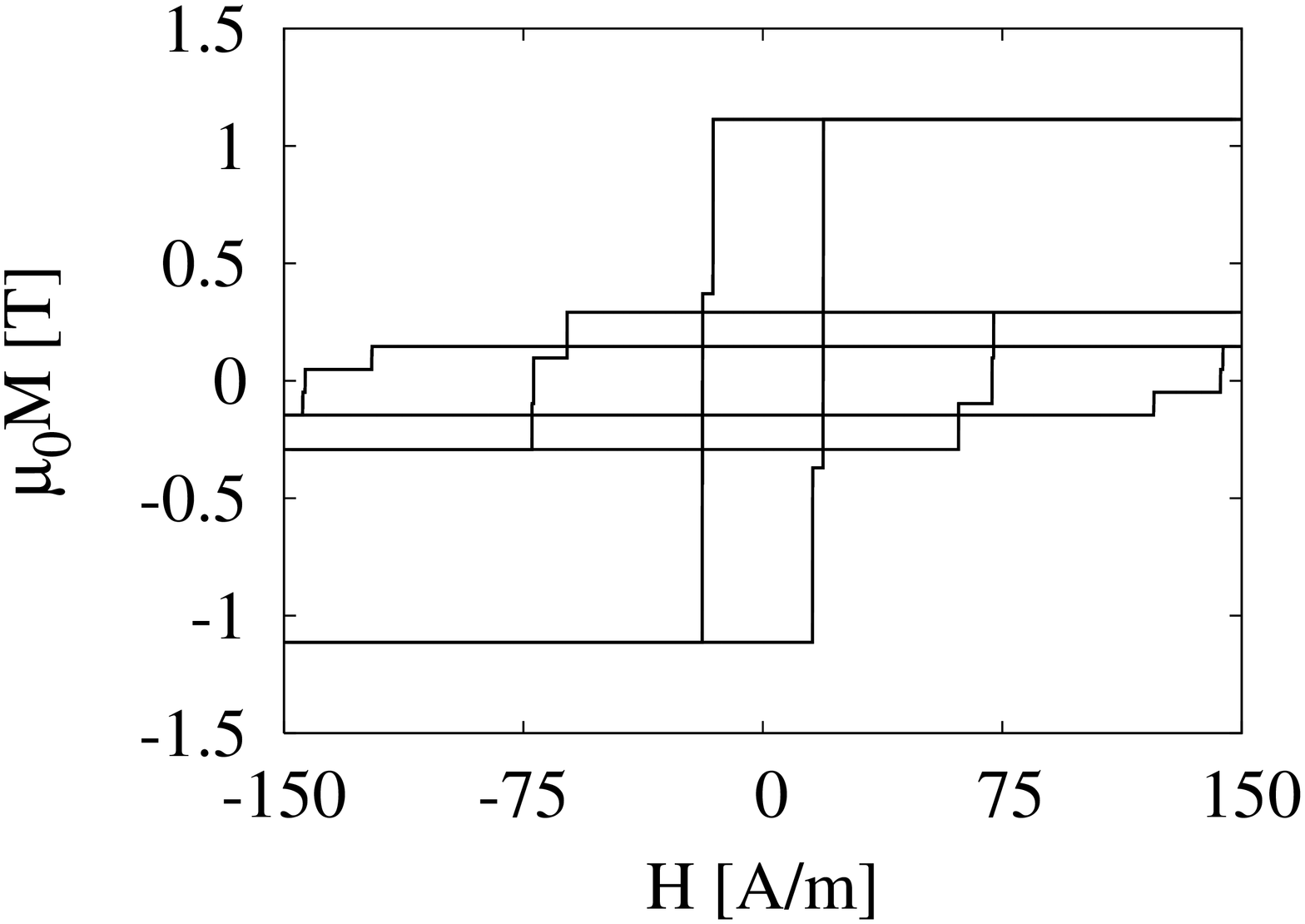}}
\end{center}
\caption{\label{fig-plot8} (a) Area where the bistability is absent. The coordinates $(x,z)$ mark the centers of the wires 
A, B, C
with respect to the center of coordinates. The line of empty squares marks the positions of the centre of wire 
A for Fig. 8 b; 
the positions of wires B and C change simultaneously and the symmetry of tetrahedron is preserved. (b) The 
horizontal part $\Delta$ of the hysteresis loop against the position of wire A. (c) Three hysteresis loops at three different
positions of the wires. As a rule, the loop is more flat when the angle between the wires is larger.}
\end{figure}

Let us assume that the magnetic field is along the OZ axis. 
The spatial configurations are: wire A along the field, wire B in the same plane XZ forms an angle $\pi /4$ (Fig. 2); 
one wire along the field, another perpendicular, also in the plane XZ but with Y
different by 2 mm (Fig. 3); two wires in the parallel planes XZ with distance 2 mm one from another, both 
wires form angles $\pm \pi /4$ with the field (Fig. 4); two wires in the same plane XZ, one perpendicular 
to the field, another forms angle $\phi$
with the former, as clock hands (Fig. 5); two wires in the same plane XZ, one parallel to the field, another forms 
angle $\phi$ with the former, as clock hands (Fig. 6); two wires in the same plane XZ form the same angle 
with the field,
rotating as bird wings (Fig. 7); three wires along three edges of a tetrahedron, form the 
same angles with the field (Fig. 8). In the case shown in Fig. 6 we include also the tensile stress, 
as large as to enhance the switching field $H*$ from 10 to 15 A/m. This is shown to change remarkably 
the area where the bistability is absent.
We also show some parameters of the obtained hysteresis loops. The width of the horizontal part of 
the loops which is responsible for the wire-wire interaction \cite{vel}, is denoted as $\Delta$. The right-shift of the 
loop center along the field axis is denoted as $s$, and the width of the loop - as $w$. 
In several cases, the results reflect obvious properties of the systems. For example, for the wires almost perpendicular
to the applied field, the hysteresis loop must be very wide. We note that when the wires are not parallel to the 
applied field, the measured and calculated $\Delta$ ceases its simple interpretation as the interaction field, but
depends also on the orientation of wires with respect to the external field. 

\section{Conclusions}

To summarize main shortcomings of our approach: {\it i)} the wire-wire interaction is approximated with an assumption
on the uniform wire magnetization, {\it ii)} our considerations are limited to the case of strict bistability, where 
the domain wall moves till the end of the wire at the same external field as it started to move. As it was remarked 
above, the former approximation can be justified if the distance between wires is much larger (at least one order of
magnitude) than the wire diameter. The second limitation excludes many cases which can be of interest. In these 
cases the dynamics of domain walls depends on the field amplitude and frequency \cite{kru}, and the numerical 
solution of the equation of motion of the domain wall seems to be the only proper method.

Even with the drawbacks listed above, the results of the model calculations point out that spatial configurations of 
bistable wires offer a rich variety of hysteresis loops. The parameters of the loops can be controlled by the 
modification of mutual positions of the wires, the angles formed by the wires and the applied magnetic field and of the
applied tensile stress. The obtained hysteresis loop reveal stable states, which remain until the external field
is enhanced above some critical values. Then, these states encode the information on previous conditions applied
to the wire. \\
\newline
{\bf Acknowledgements} K. K. and P. G. are grateful to Julian Gonz{\'a}lez and Arcady Zhukov for helpful 
discussions and kind hospitality.


\end{document}